\documentclass[superscriptaddress,showpacs,nofootinbib,aps,prd,preprint]{revtex4}
\usepackage{epsfig}
\usepackage{graphicx}
\usepackage{dcolumn}
\usepackage{bm}

\date{}
\begin{document}


\newcommand{\ds}{\displaystyle}
\newcommand{\mc}{\multicolumn}
\newcommand{\bce}{\begin{center}}
\newcommand{\ece}{\end{center}}
\newcommand{\beq}{\begin{equation}}
\newcommand{\eeq}{\end{equation}}
\newcommand{\bea}{\begin{eqnarray}}

\newcommand{\eea}{\end{eqnarray}}
\newcommand{\cont}{\nonumber\eea\bea}
\newcommand{\cl}[1]{\begin{center} {#1} \end{center}}
\newcommand{\ba}{\begin{array}}
\newcommand{\ea}{\end{array}}

\newcommand{\ab}{{\alpha\beta}}
\newcommand{\cd}{{\gamma\delta}}
\newcommand{\dc}{{\delta\gamma}}
\newcommand{\ac}{{\alpha\gamma}}
\newcommand{\bd}{{\beta\delta}}
\newcommand{\abc}{{\alpha\beta\gamma}}
\newcommand{\eps}{{\epsilon}}
\newcommand{\lam}{{\lambda}}
\newcommand{\mn}{{\mu\nu}}
\newcommand{\mpnp}{{\mu'\nu'}}
\newcommand{\Amuu}{{A_{\mu}}}
\newcommand{\Amuo}{{A^{\mu}}}
\newcommand{\Vmuu}{{V_{\mu}}}
\newcommand{\Vmuo}{{V^{\mu}}}
\newcommand{\Anuu}{{A_{\nu}}}
\newcommand{\Anuo}{{A^{\nu}}}
\newcommand{\Vnuu}{{V_{\nu}}}
\newcommand{\Vnuo}{{V^{\nu}}}
\newcommand{\Fmnu}{{F_{\mu\nu}}}
\newcommand{\Fmno}{{F^{\mu\nu}}}

\newcommand{\abcd}{{\alpha\beta\gamma\delta}}


\newcommand{\bsigma}{\mbox{\boldmath $\sigma$}}
\newcommand{\btau}{\mbox{\boldmath $\tau$}}
\newcommand{\brho}{\mbox{\boldmath $\rho$}}
\newcommand{\bpipi}{\mbox{\boldmath $\pi\pi$}}
\newcommand{\bss}{\bsigma\!\cdot\!\bsigma}
\newcommand{\btt}{\btau\!\cdot\!\btau}
\newcommand{\bnabla}{\mbox{\boldmath $\nabla$}}
\newcommand{\bphi}{\mbox{\boldmath $\tau$}}
\newcommand{\bvarphi}{\mbox{\boldmath $\rho$}}
\newcommand{\bDelta}{\mbox{\boldmath $\Delta$}}
\newcommand{\bpsi}{\mbox{\boldmath $\psi$}}
\newcommand{\bPsi}{\mbox{\boldmath $\Psi$}}
\newcommand{\bPhi}{\mbox{\boldmath $\Phi$}}
\newcommand{\bnab}{\mbox{\boldmath $\nabla$}}
\newcommand{\bpi}{\mbox{\boldmath $\pi$}}
\newcommand{\btheta}{\mbox{\boldmath $\theta$}}
\newcommand{\bkappa}{\mbox{\boldmath $\kappa$}}

\newcommand{\bA}{{\bf A}}
\newcommand{\bB}{\mbox{\boldmath $B$}}
\newcommand{\bE}{\mbox{\boldmath $E$}}
\newcommand{\bp}{\mbox{\boldmath $p$}}
\newcommand{\bk}{\mbox{\boldmath $k$}}
\newcommand{\bc}{\mbox{\boldmath $c$}}
\newcommand{\bq}{\mbox{\boldmath $q$}}
\newcommand{\bfe}{{\bf e}}
\newcommand{\bb}{\mbox{\boldmath $b$}}
\newcommand{\br}{\mbox{\boldmath $r$}}
\newcommand{\bR}{\mbox{\boldmath $R$}}

\newcommand{\fph}{${\cal F}$}
\newcommand{\aph}{${\cal A}$}
\newcommand{\dph}{${\cal D}$}
\newcommand{\fpi}{f_\pi}
\newcommand{\mpi}{m_\pi}
\newcommand{\Tr}{{\mbox{\rm Tr}}}
\def\Qb{\overline{Q}}
\newcommand{\delu}{\partial_{\mu}}
\newcommand{\delo}{\partial^{\mu}}
%
%
\newcommand{\up}{\!\uparrow}
\newcommand{\upup}{\uparrow\uparrow}
\newcommand{\updo}{\uparrow\downarrow}
\newcommand{\uu}{$\uparrow\uparrow$}
\newcommand{\ud}{$\uparrow\downarrow$}
\newcommand{\auu}{$a^{\uparrow\uparrow}$}
\newcommand{\aud}{$a^{\uparrow\downarrow}$}
\newcommand{\pu}{p\!\uparrow}

\newcommand{\qp}{quasiparticle}
\newcommand{\sa}{scattering amplitude}
\newcommand{\ph}{particle-hole}
\newcommand{\qcd}{{\it QCD}}
\newcommand{\integ}{\int\!d}
\newcommand{\ie}{{\sl i.e.~}}
\newcommand{\etal}{{\sl et al.~}}
\newcommand{\etc}{{\sl etc.~}}
\newcommand{\rhs}{{\sl rhs~}}
\newcommand{\lhs}{{\sl lhs~}}
\newcommand{\eg}{{\sl e.g.~}}
\newcommand{\ef}{\epsilon_F}
\newcommand{\sigt}{d^2\sigma/d\Omega dE}
\newcommand{\sige}{{d^2\sigma\over d\Omega dE}}
\newcommand{\rpaeq}{\beq
\left ( \begin{array}{cc}
A&B\\
-B^*&-A^*\end{array}\right )
\left ( \begin{array}{c}
X^{(\kappa})\\Y^{(\kappa)}\end{array}\right )=E_\kappa
\left ( \begin{array}{c}
X^{(\kappa})\\Y^{(\kappa)}\end{array}\right )
\eeq}
\newcommand{\ket}[1]{| {#1} \rangle}
\newcommand{\bra}[1]{\langle {#1} |}
\newcommand{\ave}[1]{\langle {#1} \rangle}
\newcommand{\half}{{1\over 2}}

\newcommand{\singlespace}{
    \renewcommand{\baselinestretch}{1}\large\normalsize}
\newcommand{\doublespace}{
    \renewcommand{\baselinestretch}{1.6}\large\normalsize}
\newcommand{\bftau}{\mbox{\boldmath $\tau$}}
\newcommand{\bfalpha}{\mbox{\boldmath $\alpha$}}
\newcommand{\bfgamma}{\mbox{\boldmath $\gamma$}}
\newcommand{\bfxi}{\mbox{\boldmath $\xi$}}
\newcommand{\bfbeta}{\mbox{\boldmath $\beta$}}
\newcommand{\bfeta}{\mbox{\boldmath $\eta$}}
\newcommand{\bfpi}{\mbox{\boldmath $\pi$}}
\newcommand{\bfphi}{\mbox{\boldmath $\phi$}}
\newcommand{\bfR}{\mbox{\boldmath ${\cal R}$}}
\newcommand{\bfL}{\mbox{\boldmath ${\cal L}$}}
\newcommand{\bfM}{\mbox{\boldmath ${\cal M}$}}
\def\dblint{\mathop{\rlap{\hbox{$\displaystyle\!\int\!\!\!\!\!\int$}}
    \hbox{$\bigcirc$}}}
\def\ut#1{$\underline{\smash{\vphantom{y}\hbox{#1}}}$}

\def\UNITY{{\bf 1\! |}}
\def\Pom{{\bf I\!P}}
\def\lsim{\mathrel{\rlap{\lower4pt\hbox{\hskip1pt$\sim$}}
    \raise1pt\hbox{$<$}}}         
\def\gsim{\mathrel{\rlap{\lower4pt\hbox{\hskip1pt$\sim$}}
    \raise1pt\hbox{$>$}}}         
\def\beq{\begin{equation}}
\def\eeq{\end{equation}}
\def\bea{\begin{eqnarray}}
\def\eea{\end{eqnarray}}

\title{Exclusive coherent production of heavy vector mesons in nucleus-nucleus collisions at LHC}

\author{A. Cisek}%
\email{Anna.Cisek@ifj.edu.pl}
\affiliation{Institute of Nuclear Physics PAN, PL-31-342 Krak\'ow, Poland}
\author{W. Sch\"afer}%
\email{Wolfgang.Schafer@ifj.edu.pl}
\affiliation{Institute of Nuclear Physics PAN, PL-31-342 Krak\'ow, Poland}
\author{A. Szczurek}%
\email{Antoni.Szczurek@ifj.edu.pl}
\affiliation{Institute of Nuclear Physics PAN, PL-31-342 Krak\'ow, Poland}
\affiliation{Institute of Physics, University of Rzesz\'ow, Poland}

\begin{abstract}
Heavy nuclei at collider energies are a source 
of high energy Weizs\"acker-Williams photons. This photon flux 
may be utilized to study high energy photon-nucleus interactions. 
Here we concentrate on the coherent diffractive production 
of heavy vector mesons on nuclear targets and show how it probes the 
unintegrated glue of the nucleus in the saturation domain.
We present predictions for rapidity distributions
of exclusive coherent $J/\Psi$ and $\Upsilon$ mesons which
can be measured by the ALICE experiment at the LHC.

\end{abstract}
\date{\today}
\maketitle


\section{Introduction}

The exclusive photo- and electroproduction of vector mesons 
at high energies has recently been thoroughly studied
at the electron--proton collider HERA (for a review, see
\cite{INS06}. 
In a Regge picture, this process is driven by a $t$-channel 
Pomeron exchange. The HERA data, which range from
low to high photon virtuality as well as from light to heavy
vector mesons, have given an intriguing insight into the
Pomeron physics from soft to hard processes.
Overall a consistent phenomenology emerges, in which
the QCD-Pomeron exchange is modelled by the gluon-ladder
exchange and quantified by the unintegrated gluon distribution of
the target proton or the color dipole-proton cross section. 

In recent years, the very-small $x$ behaviour of the gluon 
structure function has been of great interest in the context
of unitarity effects and the saturation phenomena
\cite{Saturation}.

The multiple scattering/absorption effects associated with
the gluon saturation physics are naturally enhanced on a
large nuclear target. From this point of view the 
coherent diffractive production of vector mesons on heavy nuclei is 
very interesting. 
Clearly, the best option to study the small-$x$ nuclear glue 
would be a dedicated electron-ion collider \cite{EIC}, where
for example one could measure nuclear structure functions at 
perturbatively large $Q^2$. Lacking such a facility, 
there appears to be no easy experimental access to the nuclear 
gluon distribution.

Here, the exclusive photoproduction of heavy vector mesons 
$J/\Psi,\Upsilon$ opens up a new possibility to experimentally
study the small-$x$ nuclear glue. It is the large mass
of the heavy quark which provides the hard scale and assures 
that we may use perturbation theory even in the photoproduction limit.

Presently, nucleus-nucleus collisions at $\sqrt{s_{NN}} = 2.76$ TeV
studied at the LHC offer an access to high-energy photonuclear reactions
\cite{Baur_Review}. Here, one of the nuclei will serve as a source
of Weizs\"acker-Williams photons, while the other one plays the role
of a target. Coherent diffractive processes of interest in this work
leave the target intact and deflected only by a very small angle.
Diffractive photoproduction of light vector mesons on nuclei is an old
subject \cite{Bauer}, and with some success can be addressed using
vector-meson dominance, and within a limited range of energies,
a hadronic Glauber coupled channel type of model. 
When a hard scale is present - such as a large photon virtuality, or a 
large quark mass - it proves more efficient to use perturbative QCD 
degrees of freedom. In particular the color-dipole formulation
to which we will now turn allows to account for nuclear effects in a
rather straightforward manner.    

\section{Amplitude and cross section for $\gamma A \to V A$ reaction}
\subsection{Vector meson production in the color-dipole picture}

\begin{figure}[!htp]
\begin{center}
\includegraphics[width=.7 \textwidth]{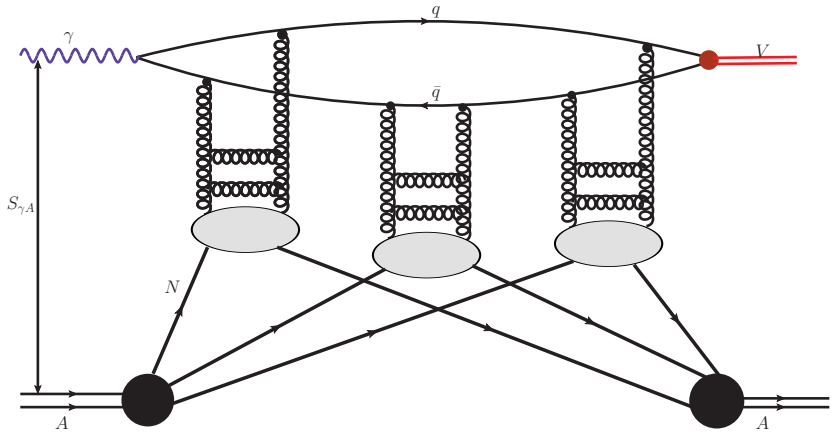}
\caption{Sample diagram for the $\gamma A \to V A$ amplitude. At $x\sim x_A$
the relevant degree of freedom is the $Q \bar Q$ dipole. 
All multipomeron exchange diagrams are summed up by the Glauber-Gribov
multiple scattering theory for color dipoles.}
\label{fig:qqbar-rescattering}
\end{center}
\end{figure}

It is useful to start from the color-dipole formulation \cite{KolyaVM} of
vector meson photoproduction. This formulation is essentially
equivalent to the $k_\perp$-factorization approach which we previously used for
the free proton target \cite{RSS08}, and to which we turn later in this work.
The multiple scattering corrections relevant for nuclear targets are however 
more easily derived in impact parameter space. 

The forward amplitude for vector-meson photoproduction on the nuclear target 
at center-of mass energy W ($x = m_V^2/W^2$), takes the form

\begin{eqnarray}
A(\gamma A \to Vp;W) &=&  i \, \bra{V} \sigma_A(x,\br) \ket{\gamma}
= 2 \, i \int d^2\bb \,  \bra{V} \Gamma_A(x,\bb,\br) \ket{\gamma}
\nonumber \\
&=& 2 \, i \, \int d^2\bb \int_0^1 dz \int d^2\br \, \psi^*_V(z,\br) \psi_\gamma(z,\br) 
\, \Gamma_A(x,\bb,\br) \, .
\label{eq:Dipole_Nucleus}
\end{eqnarray}

Here $\psi_V(z,\br),\psi_\gamma(x,\br)$ are the light-cone wave-functions 
of the vector meson and the photon, respectively. For simplicity
we suppress the summation over (anti-)quark helicities.
The explicit form of the wave function are not important for the 
argument in this section, we will later give all explicit formulas in 
momentum space.

The amplitude can be easily generalized to finite transverse 
momentum transfer $\bDelta$:
\begin{eqnarray}
A(\gamma A \to Vp;W,\bDelta) &=&
2i \int d^2\bb \, \exp[-i \bb \bDelta] \,  
\bra{V} \Gamma_A(x,\bb,\br) \ket{\gamma}
\, .
\end{eqnarray}
The differential cross section is then given by ($t = -\bDelta^2$):
\begin{eqnarray}
{d \sigma(\gamma A \to V A;W) \over dt}=  
{d \sigma(\gamma A \to V A;W) \over d\bDelta^2} 
= {1 \over 16 \pi} \Big|A(\gamma A \to Vp;W,\bDelta)
\Big|^2  \, .
\end{eqnarray}

Following \cite{NZ91}, at small $x$ ($x \lsim x_A = 0.1 A^{-1/3}$), the
multiple scatterings of the color-dipole can be summed up by a Glauber-series 
for color dipoles. The color-dipole-nucleus amplitude in impact 
parameter space for $x \sim x_A$ 
can then be given in terms of the color-dipole-proton cross section:
\begin{eqnarray}
\Gamma_A(x_A,\bb,\br) =
1 - \exp\big(-{1 \over 2} \sigma(x_A,\br) T_A(\bb) \Big) \, .
\end{eqnarray}
\label{eq:Glauber_dipole}
In practice that means $x_A \sim 0.01$. 
For smaller $x$ one must take into account
higher, $q \bar q g$-Fock states as we discuss below.

In order to quantify the size of 
nuclear (multiple scattering) effects, one often
compares to the impulse approximation. The latter works well 
if multiple scatterings
are weak, i.e. if the nuclear opacity 
$\sigma(x_A,\br) T_A(\bb)/2$ is small.
In impulse approximation we assume that only one of the nucleons in
the nucleus participates in the interaction, and all others are spectators.
Expanding the Glauber-exponential 
to the first order we obtain (hereafter 
IA stands for impulse approximation):
\begin{eqnarray}
A_{\mathrm{IA}}(\gamma A \to VA;W,\bDelta) = 
i \, \bra{V} \sigma(x,\br) \ket{\gamma} \, 
\int d^2\bb \exp[-i \bb\bDelta] T_A(\bb) \, .
\end{eqnarray}
The total cross section in impulse approximation would then be
\begin{eqnarray}
\sigma_{\mathrm{tot;IA}}(\gamma A \to V A;W) = 
4 \pi {d \sigma(\gamma p \to V p) \over dt}\Big|_{t=0} 
\, \int d^2\bb \, T_A^2(\bb) \, .
\end{eqnarray}
Let us introduce the ratio of the full nuclear cross section 
to the impulse approximation result:
\begin{eqnarray}
R_{\mathrm{coh}}(W) = {
\sigma_{\mathrm{tot}}(\gamma A \to V A;W) \over \sigma_{\mathrm{tot,IA}}(\gamma A \to V A;W) } \, .
\label{eq:R_coh}
\end{eqnarray}
Then we can also express the total photoproduction cross section
on the nucleus as:
\begin{eqnarray}
\sigma_{\mathrm{tot}}(\gamma A \to V A;W) = R_{\mathrm{coh}}(W) 
\, \cdot  4 \pi {d \sigma(\gamma p \to V p) \over dt}\Big|_{t=0} 
\, \int d^2\bb \, T_A^2(\bb) \, .
\end{eqnarray}
Here 
\begin{equation}
{d \sigma(\gamma p \to V p) \over dt}\Big|_{t=0} \approx 
B_V \cdot \sigma_{\mathrm{tot}}(\gamma p \to V p) \, ,
\end{equation}
where $B_V$ is the diffraction slope,
can be taken from experimental 
data \cite{ZEUS:JPsi,H1:JPsi,ZEUS:Upsilon}.

The integral over the nuclear optical density squared 
behaves parametrically as
(see e.g. \cite{NSS}):
\begin{eqnarray}
 \int d^2\bb \, T_A^2(\bb) = C_A \cdot {3 A^2 \over 4 \pi R_{\mathrm{ch}}^2} \, ,
\end{eqnarray}
where $R_{\mathrm{ch}}$ is the nuclear charge radius, and $C_A$ is a 
number of order unity which depends on the shape of $T_A(\bb)$. 
In the numerical calculations, we use a realistic nuclear density, 
as parametrized in \cite{Lukyanov}.

\subsection{Momentum space formulation of vector-meson production on nuclei}

We can bring the photoproduction amplitude for the nuclear target 
into the similar
$k_\perp$-factorization form as the result for the free proton. 
The only difference is that now the unintegrated
gluon distribution of the proton, will be replaced by the appropriatly defined
unintegrated glue of the nucleus, explicitly constructed in the treatment
of the diffractive $\pi A \to q\bar q A$ process in \cite{NSS}.

Recall now, that for the free-nucleon target, dipole cross 
section and unintegrated gluon 
distribution are related by \cite{NZ93}
\begin{eqnarray}
\sigma(x,\br) = \sigma_0 
\int d^2\bkappa \Big[ 1 - e^{i \bkappa \br} \Big]
\,\alpha_S f(x,\bkappa) \, ,
\end{eqnarray}
where we pulled out $\sigma_0=\sigma_0(x)$, so that $f$ is 
normalized to unity:
\begin{eqnarray}
f(x,\bkappa) ={1 \over \sigma_0} \,  {4 \pi \over N_c} \,
{1 \over \bkappa^4} \,  {\partial G_N(x,\bkappa^2) \over \partial 
\log(\bkappa^2)} \, \, , \, \, \sigma_0(x) = \int d^2\bkappa
 {4 \pi \over N_c} \,
{1 \over \bkappa^4} \,  {\partial G_N(x,\bkappa^2) \over \partial 
\log(\bkappa^2)} \, . 
\end{eqnarray}
\label{eq:small_f}
We can analogously introduce the impact-parameter dependent
unintegrated gluon distribution of the nucleus $\phi(\bb,x,\bkappa)$ 
through the relation
\begin{eqnarray}
\Gamma_A(x,\bb,\br) = \int d^2\bkappa \Big[ 1 - e^{i \bkappa \br} \Big]
\phi(\bb,x,\bkappa) \, .
\label{eq:def_Nuc_glue}
\end{eqnarray}
In terms of the nuclear glue $G_A$, the function $\phi(\bb,x,\bkappa)$ fulfills
\begin{eqnarray}
\phi(\bb,x,\bkappa) = {2 \pi \alpha_S(\bkappa) \over N_c} {1 \over \bkappa^4} \, 
{\partial G_A(\bb,x,\bkappa^2) \over \partial \log \bkappa^2 d^2\bb} \, .
\end{eqnarray}
In the forward scattering limit, i.e.
for $\bDelta =0$ the photoproduction amplitude given in \cite{INS06} 
can be brought in the form (see \cite{RSS08}):
\begin{eqnarray}
\Im m \, {\cal T}(W,\Delta^2 = 0) =
W^2 \frac{c_v \sqrt{4 \pi \alpha_{em}}}{4 \pi^2} \, 2 \, 
 \int_0^1 \frac{dz}{z(1-z)}
\int_0^\infty \pi dk^2 \psi_V(z,k^2)\\
\nonumber
\int_0^\infty
 {\pi d\kappa^2 \over \kappa^4} \alpha_S(q^2) {\cal{F}}_A(x,\kappa^2)
\Big( A_0(z,k^2) \; W_0(k^2,\kappa^2) 
     + A_1(z,k^2) \; W_1(k^2,\kappa^2)
\Big) \, ,
\end{eqnarray}
where
\begin{eqnarray}
A_0(z,k^2) &=& m_Q^2 + \frac{k^2 m_Q}{M + 2 m_Q}  \, ,
\\
A_1(z,k^2) &=& \Big[ z^2 + (1-z)^2 
    - (2z-1)^2 \frac{m_Q}{M + 2 m_Q} \Big] \, \frac{k^2}{k^2+ m_Q^{2}} \, ,
\end{eqnarray}
\begin{eqnarray}
W_0(k^2,\kappa^2) &=& 
{1 \over k^2 + m_Q^2} - {1 \over \sqrt{(k^2-m_Q^2-\kappa^2)^2 + 4 m_Q^2 k^2}}
\, , 
\\
W_1(k^2,\kappa^2) &=& 1 - { k^2 + m_Q^2 \over 2 k^2}
\Big( 1 + {k^2 - m_Q^2 - \kappa^2 \over 
\sqrt{(k^2 - m_Q^2 - \kappa^2)^2 + 4 m_Q^2 k^2 }}
\Big) \, .
\end{eqnarray}
Here $m_Q$ is the heavy quark mass, and 
\begin{equation}
M^2 = {k^2 + m_Q^2 \over z (1-z)} 
\end{equation}
is the invariant mass squared of the $Q \bar Q$ system in the final state.
The strong coupling enters at the hard scale 
$q^2 = \max (\kappa^2,k^2 + m_Q^2)$.
The light-cone wave function $\psi_V$
of the vector meson is parametrised exactly as in
\cite{INS06,RSS08}, and the Gaussian form, which proved
to lead to good agreement with experiment, is adopted.

All nuclear effects are accounted for by the substitution
\begin{eqnarray}
{\alpha_S \over \bkappa^4} {\cal{F}}_A (x,\bkappa^{2}) = 
\int d^2\bb {\alpha_S \over \bkappa^4} 
{\partial G_A(\bb,x,\bkappa^2) \over \partial \log \bkappa^2 d^2\bb} 
= {N_c \over 2 \pi} \int d^2\bb \phi(\bb,x,\bkappa) \, .
\end{eqnarray}
In this way we introduce also the impact-parameter dependent 
amplitude by the relation
\begin{eqnarray}
\Im m \, {\cal T}(\gamma A \to VA)  = \int d^2 \bb   \, 
\Im m  {\cal T}(\gamma A \to VA;\bb)  \, .
\end{eqnarray}
Its normalization is such that 
\begin{eqnarray}
{d \sigma(\gamma A \to V A;W) \over d^2\bb} = {1 \over 4} \, \Big| 
{\Im m  {\cal T}(\gamma A \to VA;\bb)   \over W^2 } \Big|^2 \, .
\end{eqnarray}

\subsection{The nuclear unintegrated glue in the Glauber regime}

Here we briefly recapitulate how to calculate the nuclear unintegrated
gluon distribution from the proton unintegrated glue.

The starting point is the definition of the nuclear unintegrated glue
(\ref{eq:def_Nuc_glue}). We are interested in the regime of $x\sim x_A$,
where the Glauber representation of the color-dipole scattering amplitude
$\Gamma(x,\bb,\br)$ is valid:
\begin{eqnarray}
\Gamma(x_A,\bb,\br) = 1 - \exp[ {1 \over 2} \sigma(x_A,\br) T_A(\bb)] \, .
\end{eqnarray}
Introducing the shorthand notation
\begin{eqnarray}
\nu(\bb) = {\frac {1}{2}} \ \alpha_{S} \ \sigma_{0}(x_A) \ T_A(\bb) \,
\, ,
\end{eqnarray}
and the the multiple convolutions
\begin{eqnarray}
f^{(j)}(x,\bkappa) = \int d^2\bkappa_1 \dots d^2\bkappa_j  
\delta^{(2)}(\bkappa - \sum_i \bkappa_i) \,   f(x,\bkappa_1) \dots f(x,\bkappa_j) 
\,  , f^{(0)}(\bkappa) \equiv \delta^{(2)}(\bkappa) \, ,
\end{eqnarray}
we obtain the expansion of the Glauber exponential 
\begin{eqnarray}
\exp[-{1 \over 2} \sigma(x_A,\br) T_A(\bb)]
&=& \sum_{k \geq 0} \, \int d^2\bkappa \, \exp[i \bkappa \br] \, w_k(\bb) \, f^{(k)}(x_A,\bkappa)
\end{eqnarray}
with the Poisson-weights
\begin{eqnarray}
w_k(\bb) = \exp[-\nu(\bb)] \, {\nu^k(\bb) \over k!} \, .
\end{eqnarray}
This gives us the expression of the nuclear unintegrated gluon distribution
as an expansion over multiple convolutions of the free nucleon glue:
\begin{eqnarray}
\phi(\bb,x,\bkappa) = \sum_{j \geq 1} \, w_j(\bb) \, f^{(j)}(x,\bkappa) \, .
\end{eqnarray}
here, the $j$-th term of the expansion is the contribution 
of the interaction of $j$ nucleons
with the color dipole. Here $f^{(j)}(x,\bkappa)$ is the 
collective unintegrated glue
of $j$ nucleons, and the weight factor 
$w_j$ gives us the probability the $j$ nucleons
overlapping at impact parameter $\bb$ take part in the interaction.

It is important to realize that the nuclear unintegrated glue 
includes the multiple scattering
corrections, it is not simply a two-gluon exchange in the 
crossed channel, as in the free
nucleon glue. To some degree however, these multiple gluon exchanges 
behave like a two-gluon exchange:
the diffractive amplitude has the same form as the two-gluon exchange amplitude on the free nucleon target.

\subsection{Small-$x$-evolution: contribution of the $Q\bar Q g$-Fock state}

\begin{figure}[!htp]
\begin{center}
\includegraphics[width=.7 \textwidth]{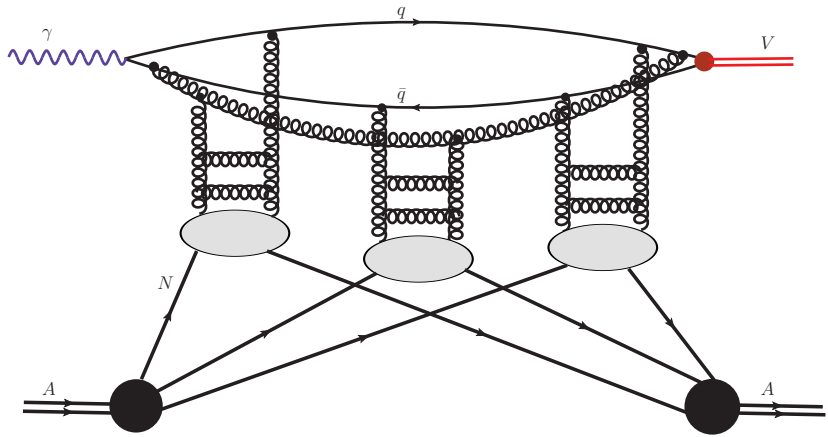}
\caption{A sample diagram which contains the multiple scattering
of a $Q \bar Q g$-Fock state. This is the first step in the nonlinear
evolution of the nuclear unintegrated glue.}
\label{rapidity}
\end{center}
\end{figure}
If we increase the $\gamma A$ center-of-mass energy successively higher
and higher $Q \bar Q g, Q\bar Q gg, \dots$ Fock-states become important.
On the free nucleon target, the effect of higher Fock-states, which
contain gluons strongly ordered in rapidity can be resummed, and
leads to the small-$x$ evolution of the then $x$-dependent dipole
cross section \cite{NZ93}. The multiple scatterings of the $Q \bar Q g$--Fock state
off a heavy nucleus are given by the first iteration
of the nonlinear Balitsky-Kovchegov \cite{BK} evolution equation
\begin{eqnarray}
{\partial \phi(\bb,x,\bp) \over \partial \log(1/x)}
= {\cal{K}}_{BFKL} \otimes  \phi (\bb,x,\bp)
 + {\cal{Q}}[\phi ](\bb,x,\bp) \, .
\nonumber 
\end{eqnarray}
The nuclear glue, which includes rescattering corrections
of $Q\bar Q$ as well as $Q\bar Qg$ Fock states is then given by
\begin{eqnarray}
\phi(\bb,x,\bp) = \phi(\bb,x_A,\bp) + \log\Big({x_A \over x} \Big) \cdot 
{\partial \phi(\bb,x,\bp) \over \partial \log(1/x)} \, .
\label{iteration}
\end{eqnarray}
In a similar manner, to obtain the ratio
$R_{\mathrm{coh}}$ of eq.(\ref{eq:R_coh}), the impulse
approximation amplitude is calculated from 
\begin{eqnarray}
\phi_{\mathrm{IA}}(\bb,x_A,\bp) = T_A(\bb) \cdot  {4 \pi \alpha_S \over N_c} \,
{1 \over \bkappa^4} 
\,  {\partial G_N(x,\bkappa^2) \over \partial \log(\bkappa^2)} \, , 
\end{eqnarray}
subject to a similar iteration (\ref{iteration}) as the full glue, 
but with the nonlinear piece omitted.
For the explicit momentum-space form of the 
infrared-regularized BK-equation, see \cite{NS06}.
A similar strategy of including the 
$Q \bar Q$ and $Q \bar Q g$-Fock states
has been followed for the nuclear structure function and inclusive
coherent diffraction in Ref.\cite{NSZZ}.
There a good agreement with available data on nuclear shadowing
has been obtained. In the numerical calculations, we use an unintegrated 
gluon distribution of the proton that has been obtained from an
analysis of HERA structure function data in \cite{IN}. 

Let us briefly discuss, how our approach differs from others available
in the literature. The first estimates of Klein and Nystrand \cite{KN} are based
on extracting an effective $J/\Psi$-nucleon cross section from photoproduction
data using vector-meson dominance ideas. They then go on to use this cross section
to evaluate the classical survival probability of mesons passing through a 
slab of nuclear matter. Goncalves and Machado \cite{GM} adopt the color dipole approach
and give a proper quantum mechanical treatment of the multiple scattering effects.
Their approach differs from ours in that they absorb all saturation effects into
the dipole-nucleon cross section which is then eikonalized. 
This is strictly speaking inconsistent with 
the nonlinear evolution of the dipole-nucleus cross section, and 
neglects the fact that multiple scatterings off different 
nucleons are enhanced by the nuclear size. 
Of course it may well be viable phenomenologically 
in a limited range of energies. 
Finally, Rebyakova et al. \cite{RSZ} use a relation of the 
diffractive amplitude to the integrated gluon distribution of the target, which
holds, with some reservations, for heavy quarks. Such an
approximation can be obtained from the $k_\perp$-factorization formalism used 
in this work to the leading logarithm in the hard scale 
(see \cite{INS06} and references therein).
There appears to be a hidden assumption that all saturation effects 
are summed up in a boundary condition of the integrated, collinear, nuclear glue.  

\section{Results and Conclusions}

\subsection{Photoproduction on nuclei: $\gamma A \to V A$}
\begin{figure}[!h]
\begin{center}
\includegraphics[width=.7\textwidth]{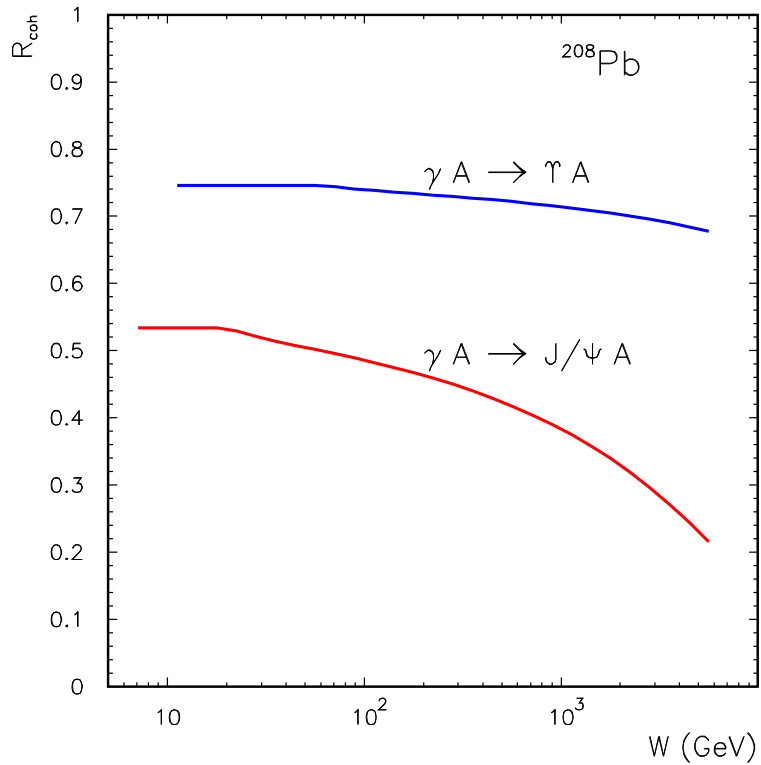}
  \caption{\label{ratio_coh}
   \small Ratio of the nuclear coherent cross section for $J/\Psi$ and $\Upsilon$ production to the Impulse Approximation cross section.}
\end{center}
\end{figure}
In Fig.\ref{ratio_coh} we show the ratio $R_{coh}(W)$ for the lead nucleus ($^{208}$Pb) for $J/\Psi$ (red line) and $\Upsilon$ (blue line) meson production. 
The deviation of $R_{coh}$ from unity 
is a measure of the strength of nuclear rescattering/absorption effects. 
We see that the nuclear effects are stronger for the 
$J/\Psi$ than for the $\Upsilon$ meson.
This is indeed to be expected, as we can most easily see in the 
dipole picture: the photoproduction amplitude probes the 
dipole cross section at the scanning radius $r_S \propto 1/m_Q$ 
\cite{KolyaVM}, and the smaller dipoles relevant for $\Upsilon$-production 
will experience smaller rescattering effects.
  
\begin{figure}[!h]
\begin{center}
\includegraphics[width=0.48\textwidth]{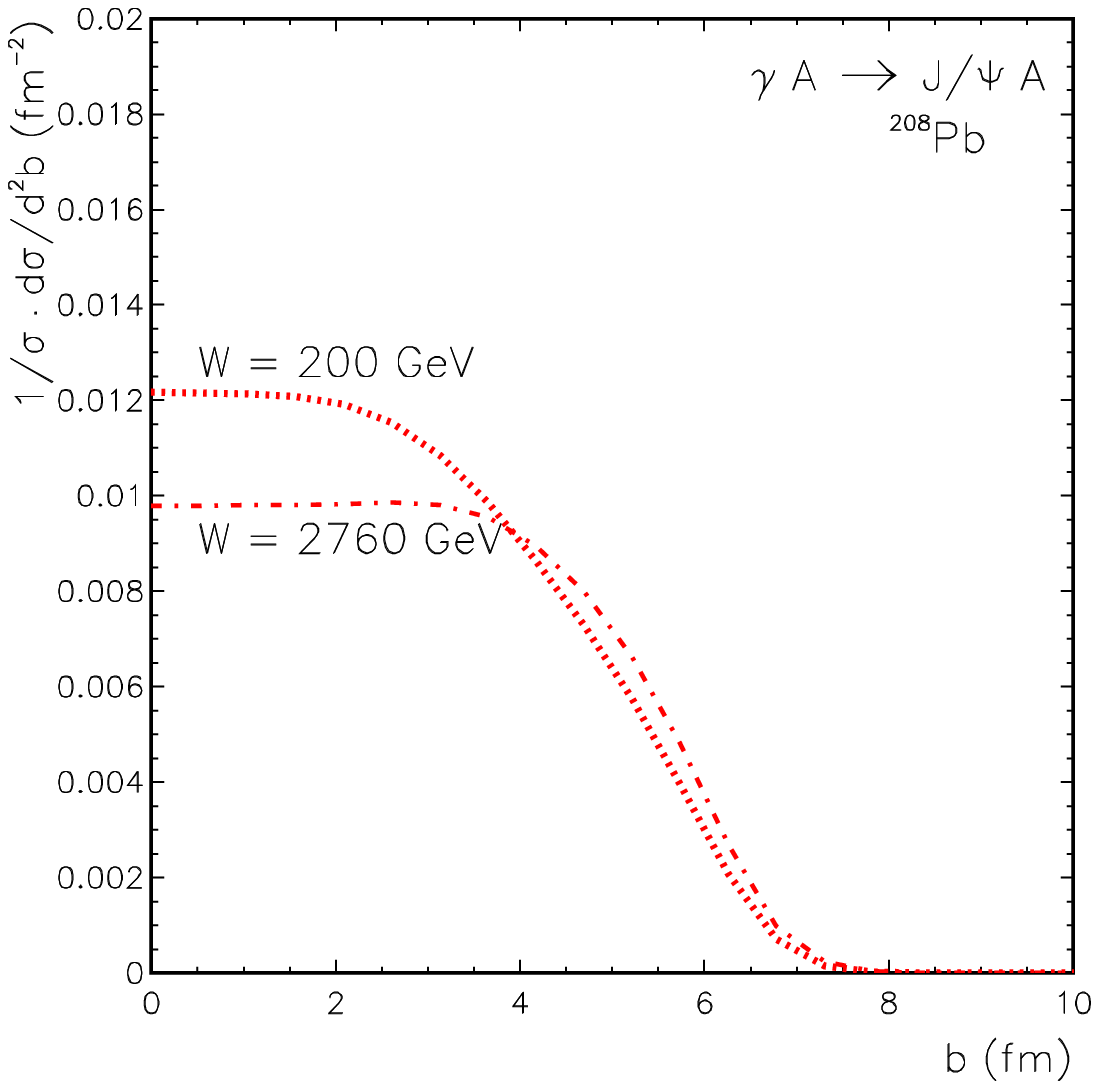}
\includegraphics[width=0.48\textwidth]{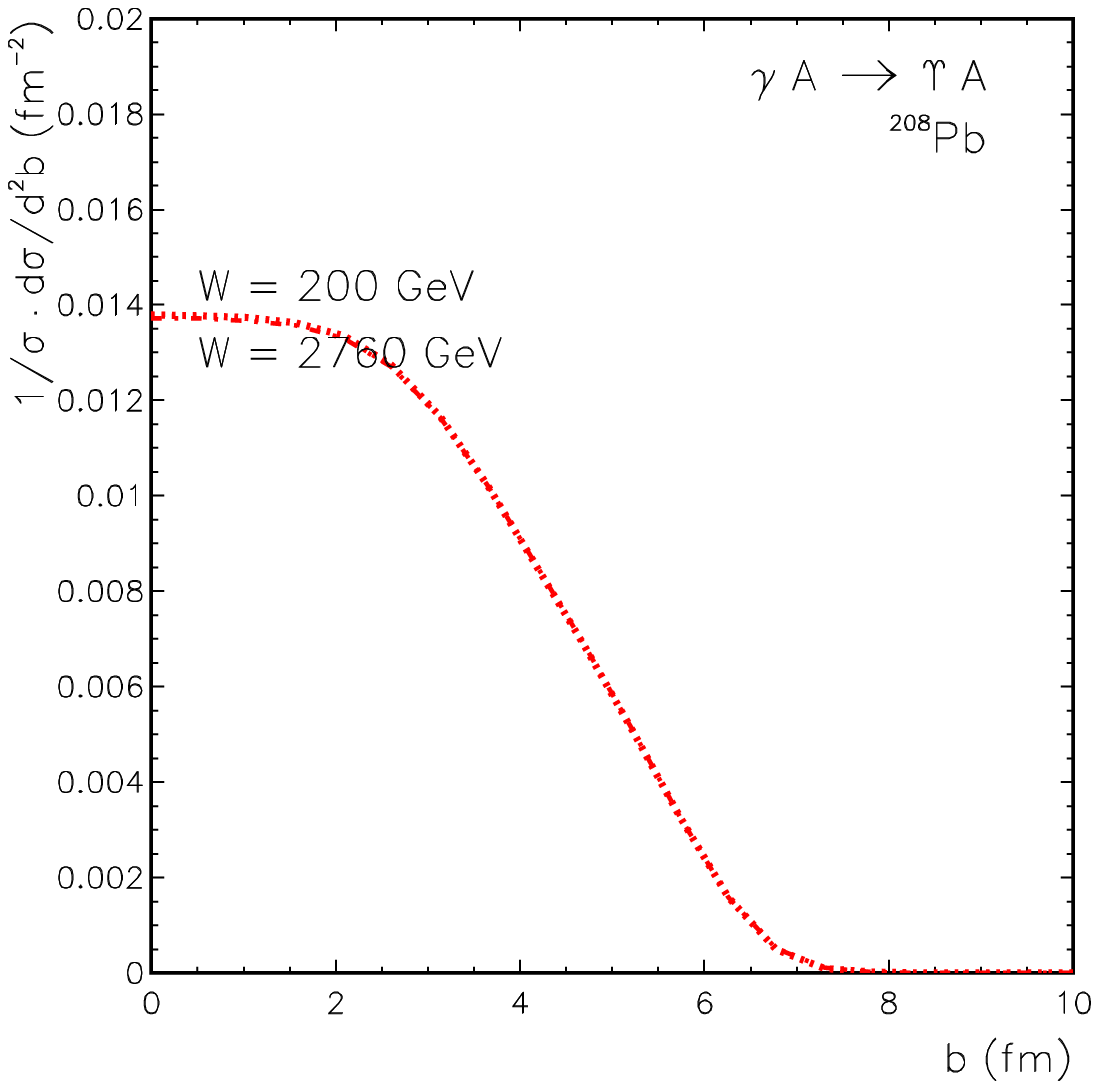}
   \caption{\label{dsig_d2b_Au_m_st}
   \small Impact parameter distributions for $J/\Psi$ and $\Upsilon$ mesons
for the $^{208}$Pb target.}
\end{center}
\end{figure}

In Fig.\ref{dsig_d2b_Au_m_st} we present the impact parameter distribution 
of vector meson photoproduction. We show results for
$J/\Psi$ and $\Upsilon$ mesons, for two different energies (W = 200, W = 2760 GeV) and lead nuclei. 

\subsection{Ultraperipheral nucleus-nucleus collisions: $AA \to AA V$}

\begin{figure}[!htp]
\begin{center}
\includegraphics[width=.45 \textwidth]{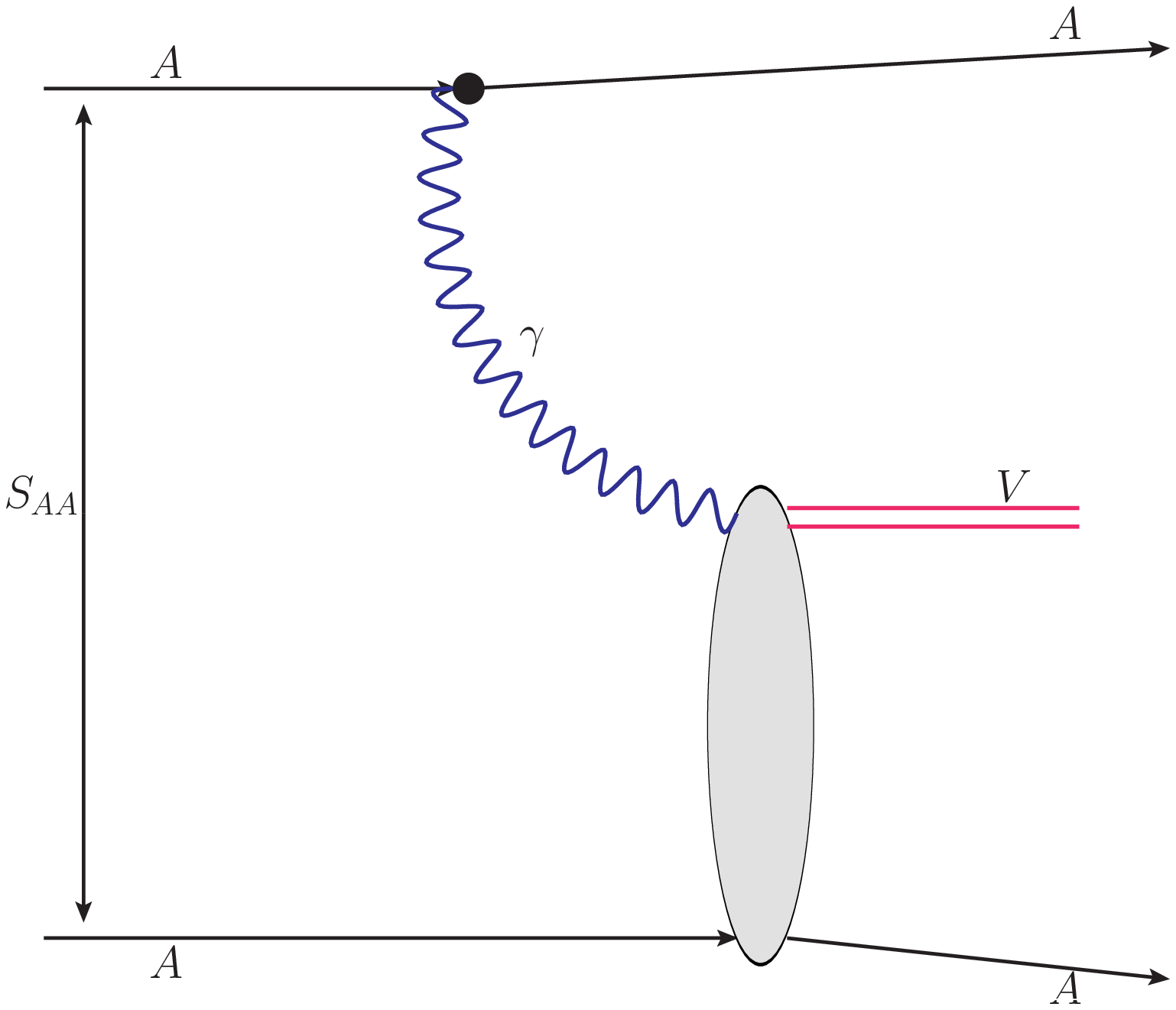}
\includegraphics[width=.45 \textwidth]{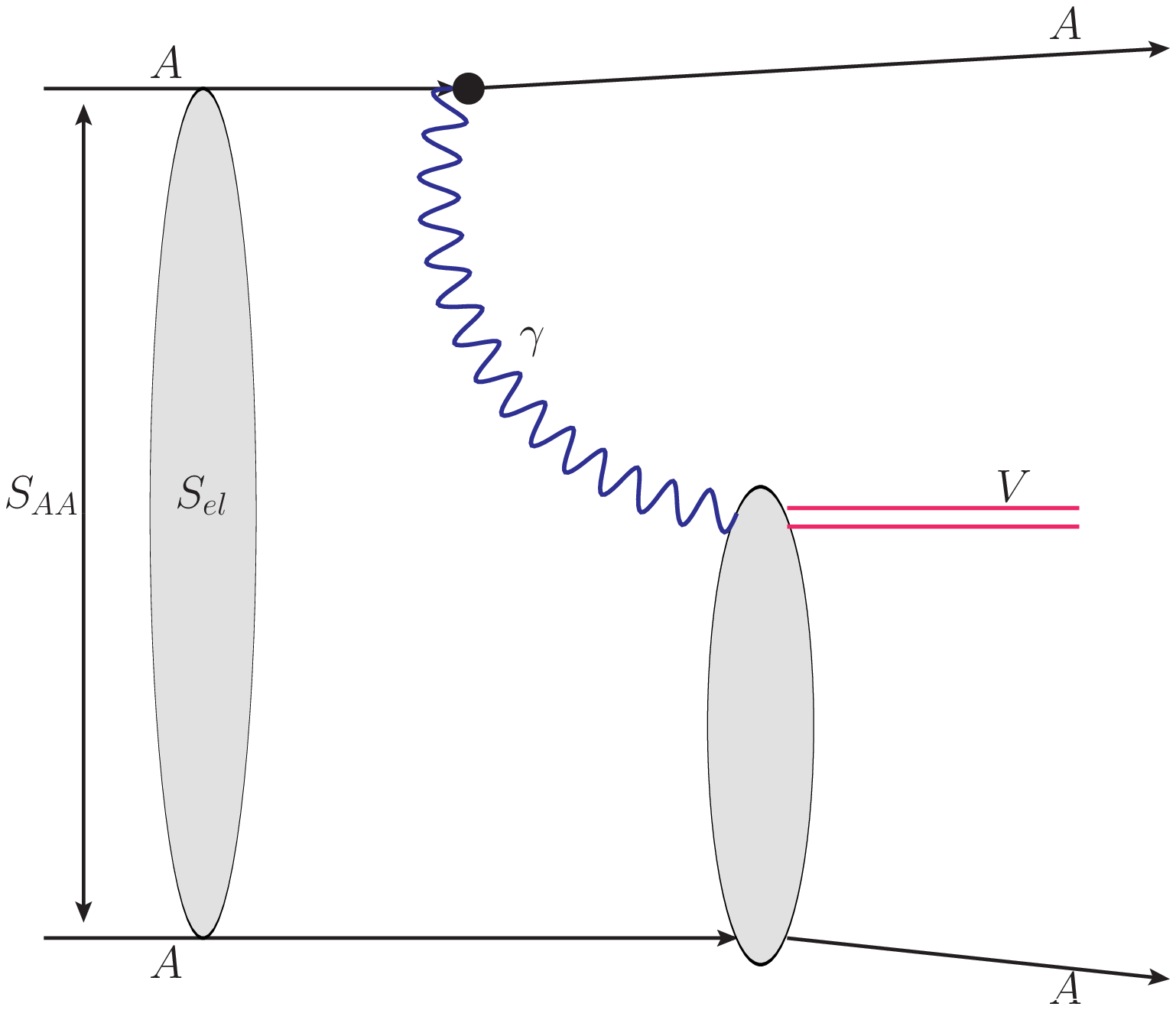}
\caption{Mechanism of exclusive vector-meson production in $AA$ collisions.
Left panel: Born diagram; Right panel: the elastic rescattering in the initial state
accounts for the unitarity effect of inelastic channels.}
\label{fig:AA_diagram}
\end{center}
\end{figure}

In the $ A A \to A A V$ processes the heavy nuclei play two different roles.
One of the nuclei is a target and the next is a source 
of high energy Weiz\"acker-Wiliams photons.
In Fig.(\ref{fig:AA_diagram}) we show Feynman diagrams for the relevant
production mechanism, the Born diagram in the left panel and a 
diagram including the absorptive correction in the right panel. 
It should be noted that one should also add the amplitude in which
the photon is emitted from the lower line. The interference of the
photon emission from the upper or lower lines in fact causes
a peculiar azimuthal correlation between the outgoing nuclei
\cite{SS07}. After integration over azimuthal angles, and at 
the Born level, the interference drops out and we can 
add the squares of both amplitudes. 
If absorptive corrections are included, a small interference 
contribution remains even after azimuthal averaging \cite{SS07}.
Below we will neglect the interference effect and evaluate
the nucleus-level cross section from the absorption corrected 
equivalent photon approximation: 
\begin{eqnarray}
\sigma(A_1 A_2 \to A_1 A_2 V;s) = \int d\omega
{d N^{\mathrm{eff}}_{A_1}(\omega) \over d\omega}
\sigma(\gamma A_2 \to V A_2; 2 \omega \sqrt{s}) 
+ (1 \leftrightarrow 2) \, .
\end{eqnarray}
To obtain the effective photon flux $dN^{\mathrm{eff}}$, one starts
from the electric field strength associated with the moving nucleus
(see for example Ref.\cite{Baur_Review} for a review and references) :
\begin{eqnarray}
\bE(\omega,\bb) = Z \sqrt{4 \pi \alpha_{em}}
\int {d^2\bq \over (2 \pi )^2} \exp[-i \bb \bq]
\ {\bq F_{em}(\bq^2 + \omega^2/\gamma^2 ) 
\over \bq^2 + \omega^2/\gamma^2 } \,.
\label{fch6_electric_field}
\end{eqnarray}
Here $F_{em}(Q^2)$ is the charge form factor of the
nucleus, $\omega$ is the photon energy, $\gamma$ the relativistic Lorentz-boost of the beam and $\bb$ is the impact parameter.
Having the electric field strength as a function of photon energy and impact
parameter we can calculate the photon flux corresponding
to the Born diagram in the left panel of Fig. \ref{fig:AA_diagram}:
\begin{eqnarray}
dN(\omega,\bb) = {d \omega \over \omega}  
{d^2 \bb \over \pi} \Big|\bE (\omega,\bb) \Big|^2 \,.
\end{eqnarray}
Finally, the effective photon flux with 
absorptive corrections included is:
\begin{eqnarray}
d N^{eff}(\omega) = \int d^2\bb \, S^2_{el}(\bb) dN(\omega,\bb) \,.
\nonumber
\end{eqnarray}
The absorptive correction $S^2_{el}(\bb)$ is shown schematically by the 
extra oval in the right panel of Fig. \ref{fig:AA_diagram}. These absorptive correction can be calculated by applying the following simple formula
(see e.g. \cite{Baur_Ferreira}):
\begin{eqnarray}
S^2_{el}(\bb) 
= \exp\Big(-\sigma_{NN} T_{A_1A_2}(\bb) \Big) \sim \theta(|\bb| - (R_1 + R_2)) \,,
\end{eqnarray} 
where $R_{1}$ and $R_{2}$ are the radii of the colliding nuclei. We remove those
configurations in the impact parameter space, when the nuclei overlap,
which at high energy means automatically their break up. Therefore
absorptive corrections have a meaning of the gap survival probability.
\begin{figure}[!h]
\begin{center}
\includegraphics[width=0.8\textwidth]{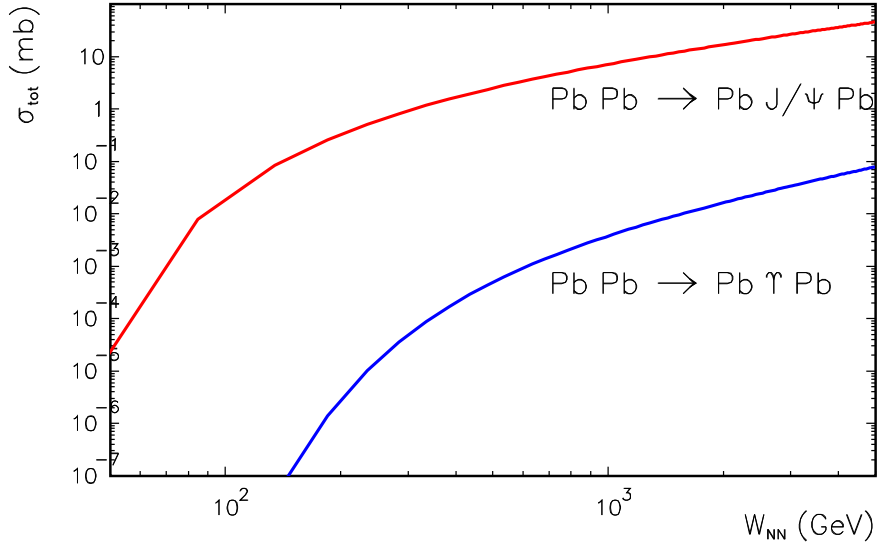}
   \caption{\label{w_tot_au}
   \small Total cross section for $Pb\ Pb \to Pb\ J/\Psi \ Pb$ and $Pb\ Pb 
\to Pb\ \Upsilon \ Pb$ as a function of nucleon-nucleon energy.}
\end{center}
\end{figure}
In Fig. \ref{w_tot_au} we present the total cross section for 
$J/\Psi$ and $\Upsilon$ production
in nucleus-nucleus collisions as a function of energy for the $^{208}Pb$ target.
The cross section for $J/\Psi$ production increases by two-orders of magnitude
when going from RHIC to LHC energy. For the $\Upsilon$ meson the corresponding
increase of the cross section is substantially bigger.
\begin{figure}[!h]
\begin{center}
\includegraphics[width=0.8\textwidth]{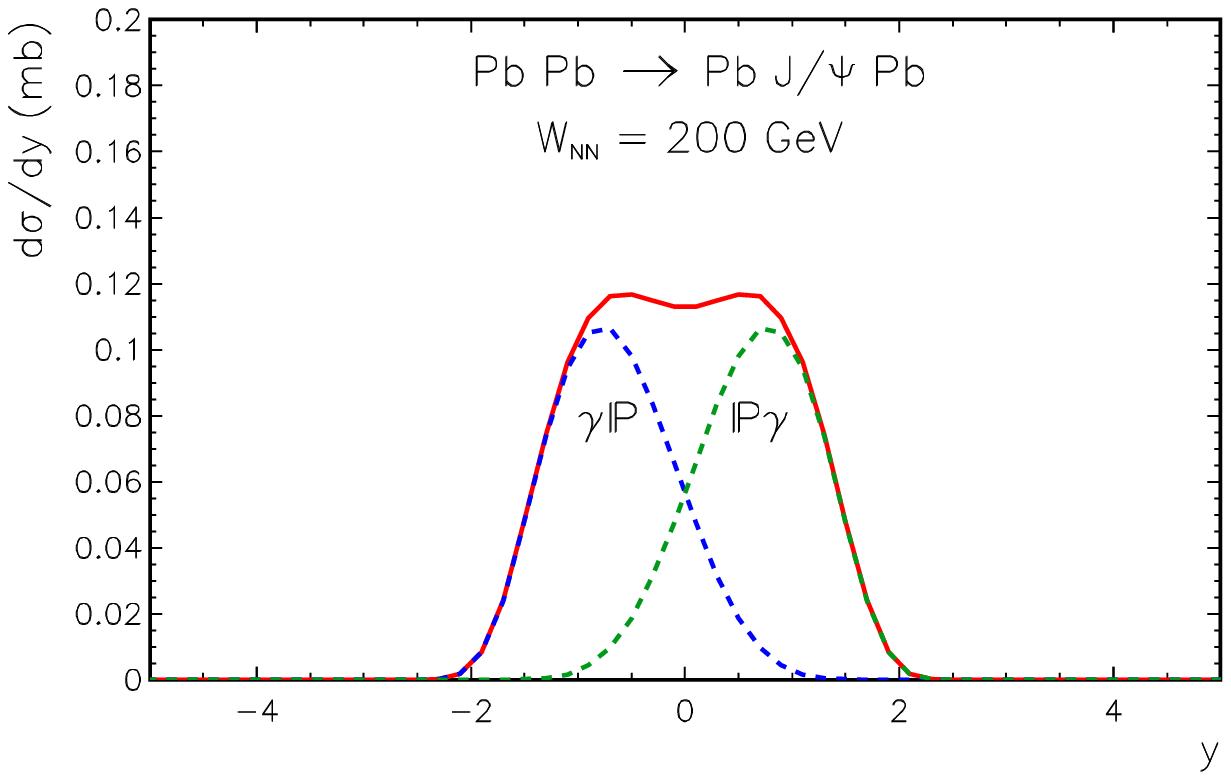}
\includegraphics[width=0.8\textwidth]{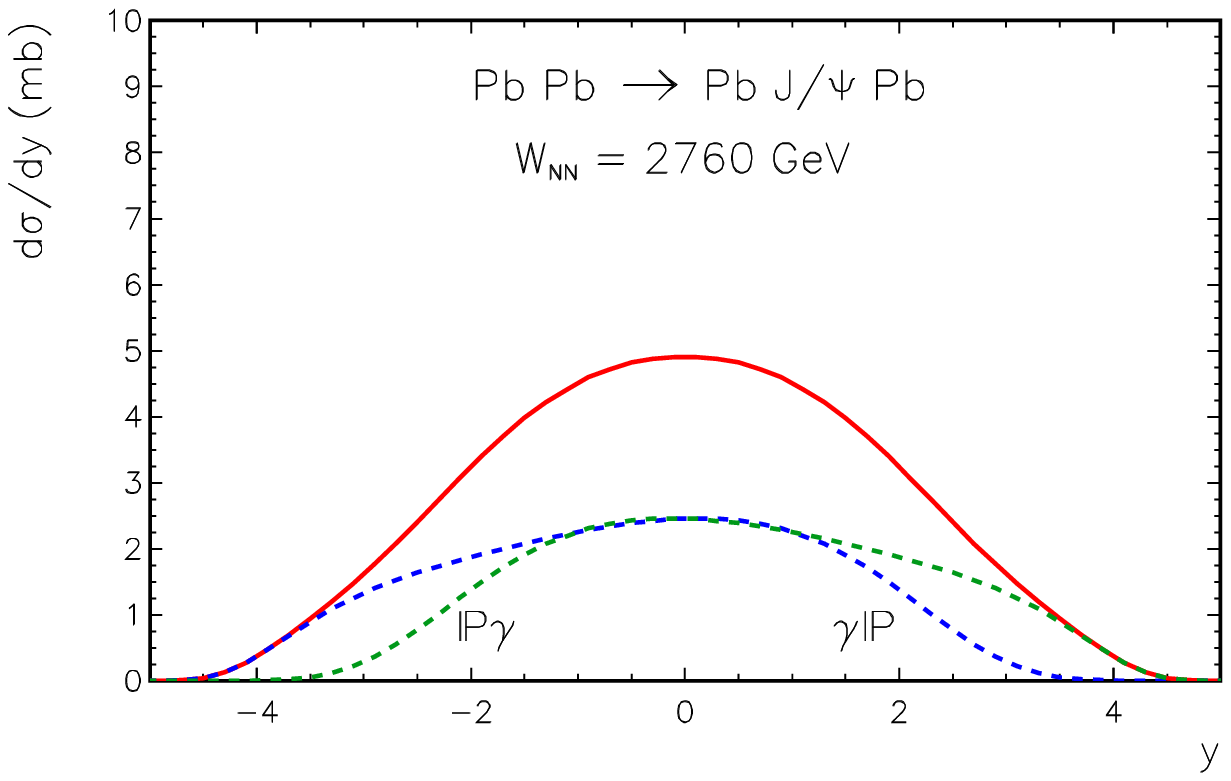}
   \caption{\label{dsig_dy_jpsi}
   \small Rapidity distribution of $J/\Psi$ for symmetric collisions of 
          lead nuclei for W$_{NN}$ = 200 GeV (upper)
          and W$_{NN}$ = 2760 GeV (lower). Individual contributions are shown separately.}
\end{center}
\end{figure}
\begin{figure}[!h]
\begin{center}
\includegraphics[width=0.8\textwidth]{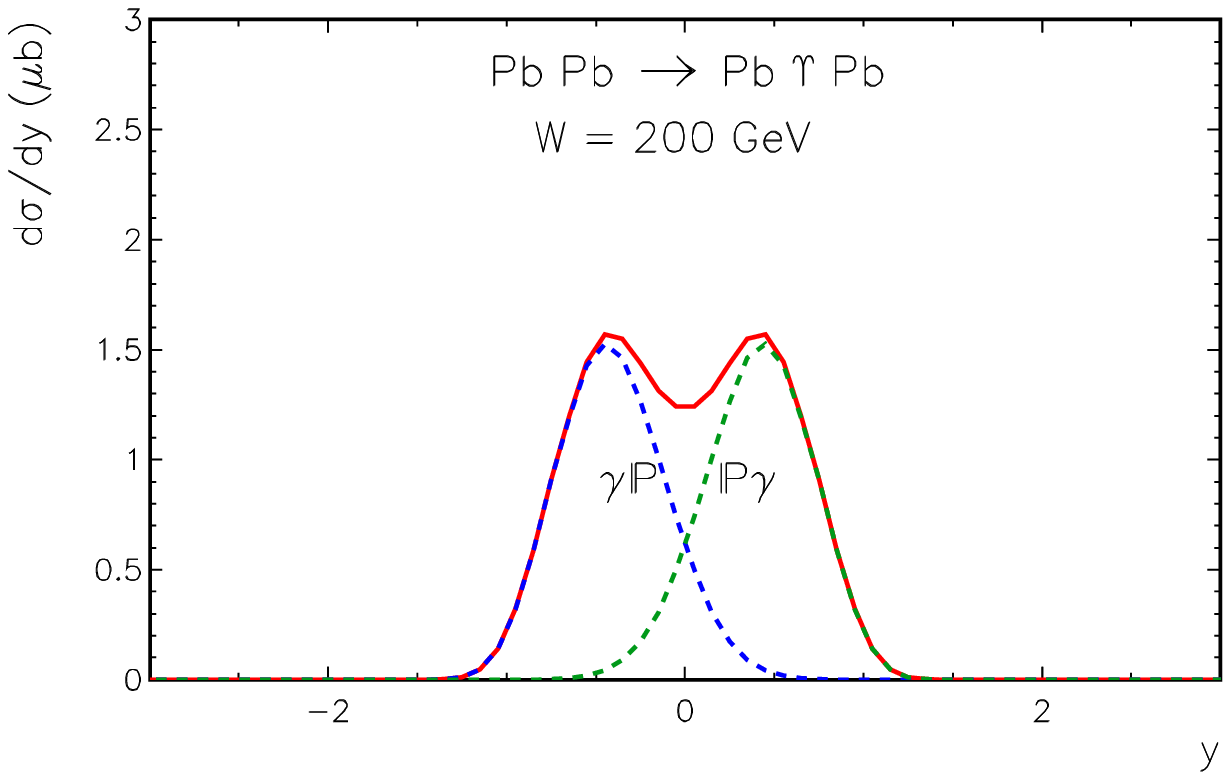}
\includegraphics[width=0.8\textwidth]{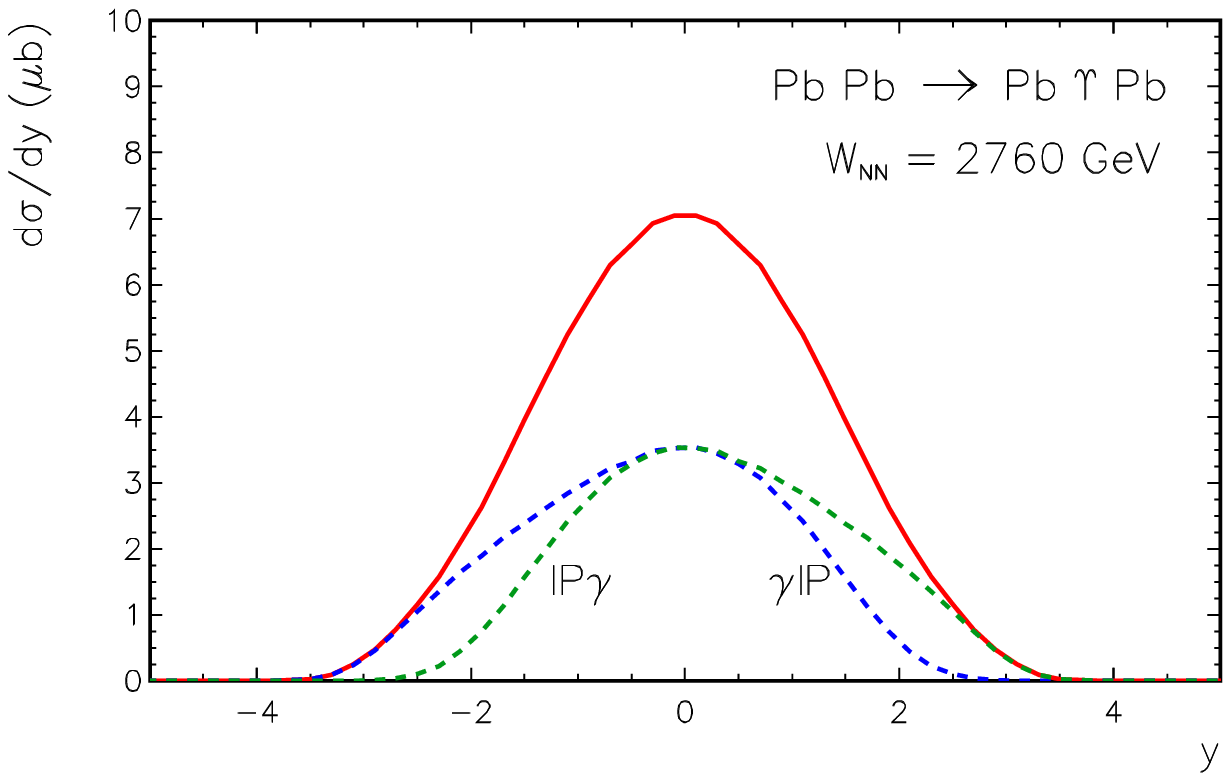}
   \caption{\label{dsig_dy_ups}
   \small Rapidity distribution of $\Upsilon$ for symmetric collisions of lead nuclei 
	for W$_{NN}$ = 200 GeV (upper) and W$_{NN}$ = 2760 GeV (lower).
          Individual contributions are shown separately.}
\end{center}
\end{figure}
In Figs. \ref{dsig_dy_jpsi} - \ref{dsig_dy_ups} we show the 
differential cross section in rapidity for the exclusive coherent 
productionof $J/\Psi$ and $\Upsilon$ mesons, in lead-lead collisions. 
The shape of the distributions strongly depends on the collision energy.
The distributions in rapidity for AA collision are much narrower 
than similar distributions in proton-proton 
collisions \cite{SS07,RSS08}.
This has its origin in the large nuclear size: the charge form factor 
$F(q^2)$ of the nucleus is much sharper compared to the proton's 
Dirac form factor $F_{1}(q^2)$. The spectrum of Weizs\"acker-Williams
photons in a proton is considerably harder than in the nucleus,
see e.g. \cite{Baur_Review}.

In summary, we presented predictions for the exclusive coherent diffractive
production of $J/\Psi$ and $\Upsilon$-mesons in collisions of heavy 
nuclei at LHC energies. Our framework takes into account not only 
the Glauber-type rescattering of color dipoles in the nuclear matter, 
but also the gluon-fusion/shadowing corrections associated with 
the rescattering of the $Q \bar Q g$-Fock-state. The predicted rapidity
distributions of mesons may be tested by the ALICE experiment at LHC.

\subsection{Acknowledgements}
W.S. would like to thank Kolya Nikolaev for numerous discussions in past years 
related to the subject of this work.
This work was supported by the MNiSW grant DEC-2011/01/B/ST2/04535.


\begin{thebibliography}{299}

\bibitem{INS06}
 I.~P.~Ivanov, N.~N.~Nikolaev and A.~A.~Savin,
  Phys.\ Part.\ Nucl.\  {\bf 37}, 1 (2006).

\bibitem{Saturation}
  K.~J.~Golec-Biernat,
  Acta Phys.\ Polon.\  {\bf B35 } (2004)  3103-3114;
  L.~McLerran,
  [arXiv:0812.4989 [hep-ph]];
  Y.~V.~Kovchegov,
  Nucl.\ Phys.\ A {\bf 854}, 3 (2011).

\bibitem{EIC} 
  A.~Deshpande, R.~Milner, R.~Venugopalan and W.~Vogelsang,
  Ann.\ Rev.\ Nucl.\ Part.\ Sci.\  {\bf 55}, 165 (2005).

\bibitem{Baur_Review} 
  G.~Baur, K.~Hencken, D.~Trautmann, S.~Sadovsky and Y.~Kharlov,
  Phys.\ Rept.\  {\bf 364}, 359 (2002).

\bibitem{Bauer} 
  T.~H.~Bauer, R.~D.~Spital, D.~R.~Yennie and F.~M.~Pipkin,
  Rev.\ Mod.\ Phys.\  {\bf 50}, 261 (1978)
  [Erratum-ibid.\  {\bf 51}, 407 (1979)].

\bibitem{KolyaVM} 
  N.~N.~Nikolaev,
  Comments Nucl.\ Part.\ Phys.\  {\bf 21}, 41 (1992);
  B.~Z.~Kopeliovich, J.~Nemchick, N.~N.~Nikolaev and B.~G.~Zakharov,
  Phys.\ Lett.\ B {\bf 309}, 179 (1993);   J.~Nemchik, N.~N.~Nikolaev, E.~Predazzi and B.~G.~Zakharov,
  Phys.\ Lett.\ B {\bf 374}, 199 (1996); Z.\ Phys.\ C {\bf 75}, 71 (1997).

\bibitem{RSS08}
  A.~Rybarska, W.~Sch\"afer and A.~Szczurek,
  Phys.\ Lett.\  B {\bf 668} 126 (2008).

\bibitem{NZ91}
  N.~N.~Nikolaev, B.~G.~Zakharov,
  Z.\ Phys.\  {\bf C49}, 607 (1991).

\bibitem{ZEUS:JPsi} 
  S.~Chekanov {\it et al.}  [ZEUS Collaboration],
  Eur.\ Phys.\ J.\ C {\bf 24}, 345 (2002).

\bibitem{H1:JPsi} 
  A.~Aktas {\it et al.}  [H1 Collaboration],
  Eur.\ Phys.\ J.\ C {\bf 46}, 585 (2006).

\bibitem{ZEUS:Upsilon} 
  J.~Breitweg {\it et al.}  [ZEUS Collaboration],
  Phys.\ Lett.\ B {\bf 437}, 432 (1998);
  S.~Chekanov {\it et al.}  [ZEUS Collaboration],
  Phys.\ Lett.\ B {\bf 680}, 4 (2009).


\bibitem{NSS}
  N.~N.~Nikolaev, W.~Sch\"afer, G.~Schwiete,
  Phys.\ Rev.\  {\bf D63 } 014020 (2001).

\bibitem{Lukyanov} 
  V.~Lukyanov, E.~Zemlyanaya and B.~Slowinski,
  [nucl-th/0308079];
V.~Lukyanov and E.~Zemlyanaya J. Phys. G {\bf 26} 357 (2000).

\bibitem{NZ93}
  N.~N.~Nikolaev, B.~G.~Zakharov,
  Z.\ Phys.\  {\bf C64 } 631 (1994).

\bibitem{BK}
  I.~Balitsky,
  Nucl.\ Phys.\  {\bf B463 } 99 (1996);
  Y.~V.~Kovchegov,
  Phys.\ Rev.\  {\bf D60 } 034008 (1999).

\bibitem{NS06}
  N.~N.~Nikolaev, W.~Sch\"afer,
  Phys.\ Rev.\  {\bf D74 } 014023 (2006).


\bibitem{NSZZ}
  N.~N.~Nikolaev, W.~Sch\"afer, B.~G.~Zakharov, V.~R.~Zoller,
  JETP Lett.\  {\bf 84 } 537 (2007).

\bibitem{IN} 
  I.~P.~Ivanov and N.~N.~Nikolaev,
  Phys.\ Rev.\ D {\bf 65}, 054004 (2002).

\bibitem{KN} 
  S.~Klein and J.~Nystrand,
  Phys.\ Rev.\ C {\bf 60}, 014903 (1999).

\bibitem{GM} 
  V.~P.~Goncalves and M.~V.~T.~Machado,
  Phys.\ Rev.\ C {\bf 84}, 011902 (2011).

\bibitem{RSZ} 
  V.~Rebyakova, M.~Strikman and M.~Zhalov,
  Phys.\ Lett.\ B {\bf 710}, 647 (2012).

\bibitem{SS07}
  W.~Sch\"afer and A.~Szczurek,
  Phys.\ Rev.\  {\bf D76}, 094014 (2007).

\bibitem{Baur_Ferreira} 
  G.~Baur and L.~G.~Ferreira Filho,
  Nucl.\ Phys.\ A {\bf 518}, 786 (1990).
\end{thebibliography}
\end{document}